\documentclass[a4paper,12pt,one column]{article}
\usepackage{graphicx}
\usepackage{float}
\usepackage{amsmath}
\usepackage{amssymb}
\usepackage{natbib}
\usepackage{hyperref}
\usepackage{color}

\title{{Undersampled dynamic X-ray tomography with dimension reduction Kalman filter}}
\author{Janne Hakkarainen$^{1,2}$, Zenith Purisha$^{1,3}$, Antti Solonen$^{4,5}$, and Samuli Siltanen$^1$}

\date{% 
   \small
    $^1$Department of Mathematics and Statistics, University of Helsinki, Finland\\
    $^2$Finnish Meteorological Institute, Helsinki, Finland\\
    $^3$School of Electrical Engineering, Aalto University, Espoo, Finland\\
    $^4$Eniram Ltd., Helsinki, Finland\\
    $^5$School of Engineering Science, Lappeenranta University of Technology, Lappeenranta, Finland\\%\\[2ex]%
    May 2018
}

\begin{document}
\maketitle

\abstract{In this paper, we consider prior-based dimension reduction Kalman filter for undersampled dynamic X-ray tomography. With this method, the X-ray reconstructions are parameterized by a low-dimensional basis. Thus, the proposed method is a) computationally very light; and b) extremely robust as all the computations can be done explicitly. With real and simulated measurement data, we show that the method provides accurate reconstructions even with very limited number of angular directions.}

\vspace{2pc}
 Submitted to \textit{IEEE Transactions on Computational Imaging.}

\section{Introduction}
Computed tomography (CT) is a non invasive method that is widely used in many applications to reveal inner structure of an unknown target using propagation of X-ray through the target from multiple view or projections. To reconstruct the attenuation values of the target, standard methods such as filtered backprojection (FBP) or Feldmann-David-Kress (FDK) have been well implemented \cite{buzug2008computed,natterer1986mathematics,natterer2001mathematical,avinash1988principles}. However, these methods are well understood only when the target is not moving (static) and dense X-ray measured data from full angles are available. 

In recent years, dynamic X-ray tomography, in which the target is non-static or changing, has received particular attention in CT, and several ideas have been proposed lately, for example \cite{bonnet2003dynamic,Roux2004,Katsevich2010,hahn2014reconstruction,hahn2014efficient,hahn2015dynamic,bubba2017shearlet,burger2017variational,niemi2015dynamic}. The goal here is to reconstruct the target in space and time from severely undersampled measurements. 

We assume that the data measured at any single time instance is too scarce for reconstructing the target properly. Our motivation stems from applications where the radiation dose needs to be minimized, fast acquisition is required, big datasets need to be avoided, or the imaging geometry is restricted. Also, patient movement during a medical CT scan results in a dynamic dataset.

We are also interested in multi-source arrangements where at each time step we have measurements only from a few angular directions. Such arrangements have received so far only little attention in the literature \cite{berninger1980multiple,robb1983high,stiel1993digital,liu2001half}; perhaps so because filtered back-projection type methods are not well-suited for the resulting sparse datasets. However, progress in the mathematics of inverse problems now allows novel reconstruction approaches.

In this paper we propose the dimension reduction Kalman filter for undersampled dynamic X-ray tomography. We demonstrate the method with both simulated and real measurement data. In comparison to many other methods, ours is computationally very light, and can compute the solution \textit{online} as the measurement takes place without the need of solving all the reconstructions at the same time. The dimension reduction Kalman filter is also based on direct computations, i.e., the solution is not obtained iteratively.  

We also show how the dimension reduction can be applied in Rauch--Tung--Striebel (RTS) smoother that can be used for post-processing the Kalman filter results. RTS smoothing improves the quality of the reconstructions in particular in the beginning of the simulation. 

This new dimension reduction version of the Kalman filter was recently introduced by \cite{solonen2016} and has been motivated and tested for (chaotic) geophysical applications, where it performed well against the standard methods like extended Kalman filter and ensemble Kalman filter. In this paper, we show that it can be used also for dynamic X-ray tomography where the application of the standard version of the Kalman filter would be too memory-consuming.

The rest of this paper is organized as follows. Section~\ref{methods} introduces measurement principle and dimension reduction for static and dynamic X-ray tomography. Section~\ref{numer} presents numerical results using both real and simulated data, and finally, Section~\ref{conclusions} concludes the paper.

\section{Materials and methods\label{methods}}

\subsection{Dynamic X-ray tomography}
In X-ray tomographic imaging, the aim is to collect information about an unknown attenuation function $x$ using available measurement data as a collection of line integrals through it. Mathematically, the Radon transform $\mathcal{H}$ represents the measurement data $y$ \cite{buzug2008computed,natterer2001mathematical,SiltanenBook}, so the problem can be modeled as
\begin{equation}
y = \mathcal{H}(x).
\end{equation}
In dynamic X-ray tomography, the attenuation function $x$ is also a function of the time $k \in \mathbb{N}$.

Using a discrete notation, the model at time step $k$ can be written as
\begin{equation}
\vec{y}_k = \mathbf{H}_k \vec{x}_k + \vec{\varepsilon}_k.
\label{ip}
\end{equation}
where $\vec{y}_k$ is the measurement vector called sinogram, $\vec{x}_k$ is the vectorized X-ray image, $ \mathbf{H}_k$ is the measurement matrix  and $\vec{\varepsilon}_k$ is the noise vector.

The aim of the static X-ray tomography, is to construct an image $\vec{x}_k$ so that the distance between $\vec{y}_k$ and $\mathbf{H}_k \vec{x}_k$ is minimized in some statistical sense. Especially when the measurement data is scarce, additional \textit{a priori} information needs to be added in the reconstruction process as there are many images $\vec{x}_k$ giving close to minimal distance between $\vec{y}_k$ and $\mathbf{H}_k \vec{x}_k$.

In the Bayesian framework, the inverse problem is recast as a stable statistical inference problem. Both the unknown image and the measurement are modelled as random vectors. Measurement information is included in the process via \textit{likelihood model}, and all prior information we have available on the unknown is built into a \textit{prior model}. These two models, or probability distributions, are combined into a \textit{posterior distribution} using the Bayes formula. Finally one tries to find the \textit{maximum a posteriori} estimate, i.e., an image that maximize the posterior distribution $p( \vec{x}_k\vert  \vec{y}_k)$ given the sinogram.

In dynamic  X-ray tomography,  the measurement function $\mathbf{H}_k$ can be substantially different at every time step, because one often wants to work in the situation where only few angular directions are available at each time step $k$, and the quality of the reconstruction is simply not good enough using the measurements from the same time step only. In dynamic systems one is not limited to finding solution $p( \vec{x}_k\vert  \vec{y}_k)$, but rather tries to find an \textit{online} (filtering) solution $p( \vec{x}_k\vert  \vec{y}_{1:k})$ using all the available observations $ \vec{y}_{1:k}$ at time $k$, or an \textit{offline} (smoothing) solution $p( \vec{x}_k\vert  \vec{y}_{1:K})$ where one first collect all the observations $ \vec{y}_{1:K}$ until the final time index $K$.

Thus, in dynamic X-ray tomography the idea is to reconstruct a set of changing X-ray images in space and time. Let us first, however, look how we can solve this inverse problem in static case.

\subsection{Prior-based dimension reduction}

There are several ways to solve the static version of the linear inverse problem (\ref{ip}) (see, e.g., the textbook \cite{SiltanenBook} for introduction). Here we discuss two ways that  will later serve us in numerical computations.

In Tikhonov regularization one finds the minimum of the following functional
\begin{equation}
J(\vec{x}_k) = \Vert  \vec{y}_k - \mathbf{H}_k \vec{x}_k \Vert^2 + \Vert \mathbf{\Gamma} \vec{x}_k \Vert^2,
\end{equation}
where $\mathbf{\Gamma}$ is the chosen Tikhonov matrix. The solution $\vec{x}_k^{\mathrm{tik}}$ can obtained from the formula
\begin{equation}
\vec{x}_k^{\mathrm{tik}} = (\mathbf{H}_k^{\mathrm T}\mathbf{H}_k + \mathbf{\Gamma}^{\mathrm T}\mathbf{\Gamma})^{-1}\mathbf{H}^{\mathrm T}_k\vec{y}_k.
\label{Tik1}
\end{equation}

In Bayesian framework one assumes that we have a prior probability distribution for our quantities of interest. Once the measurements are taken, the prior distribution can be updated, and this new updated distribution is called the posterior. If we assume Gaussian prior $\vec{x_k} \sim N(\vec{\mu}_k,\mathbf{\Sigma})$ and observation error $\vec{\varepsilon}_k \sim N(\vec{0},\mathbf{R}_k)$,  then the posterior density can be written as
\begin{equation}
p( \vec{x}_k\vert \vec{y}_k ) \propto \exp \left( -\frac{1}{2} \left(    \Vert  \vec{y}_k -\mathbf{H}_k \vec{x}_k \Vert^2_{\mathbf{R}_k} + \Vert \vec{x}_k- \vec{\mu}_k  \Vert^2_{\mathbf{\Sigma}}    \right)\right) .
\end{equation}
Now in  this linear case also the posterior is Gaussian and the mean and its error covariance matrix  can be obtained from
\begin{eqnarray}
\label{Gauss1}
\vec{x}_k^{\mathrm{est}} & = & \vec{\mu}_k + \mathbf{C}_k^{\mathrm{est}}\mathbf{H}_k^{\mathrm T}\mathbf{R}_k^{-1}(\vec{y}_k-\mathbf{H}_k \vec{\mu}_k), \\ 
\mathbf{C}_k^{\mathrm{est}} & = & (\mathbf{H}_k^{\mathrm T}\mathbf{R}_k^{-1}\mathbf{H}_k + \mathbf{\Sigma}^{-1})^{-1}.
\label{Gauss2}
\end{eqnarray}

{\bf Remark.}  Even for low resolution X-ray images of the size $N\times N$, the direct usage of the Bayes and Tikhonov matrix formulas is cumbersome as one would need to invert a matrix of the size $N^2\times N^2$. Next we show how this problem can be circumvented by using prior-based dimension reduction. 

In prior-based dimension reduction (see, e.g., \cite{Marzouk20091862}) the idea is to constrain the problem onto a subspace that contains most of the variability allowed by the prior. In recent paper \cite{solonen2016}, this concept was extended to Kalman filtering.

In dimension reduction, we start by parameterizing the state
\begin{equation}
\vec{x}_k = \vec{\mu}_k + \mathbf{P} \vec{\alpha}_k,
\end{equation}
using the prior mean $\vec{\mu}_k$ and the projection matrix  $\mathbf{P}$ that projects parameters $\vec{\alpha}_k$ from the low dimensional subspace onto the full state space.

The aim is now find the posterior distribution for the parameters  $\vec{\alpha}_k$:
\begin{equation}
p( \vec{\alpha}_k\vert \vec{y}_k )  \propto \exp \left( -\frac{1}{2} \left(    \Vert  \vec{y}_k -\mathbf{H}_k (\vec{\mu}_k + \mathbf{P} \vec{\alpha}_k) \Vert^2_{\mathbf{R}_k} + \Vert \mathbf{P} \vec{\alpha}_k  \Vert _{\mathbf{\Sigma}}  \right)\right) .
\label{P1}
\end{equation}

In prior-based dimension reduction, $\mathbf{P}$ is constructed from the prior covariance matrix $\mathbf{\Sigma}$ using singular value decomposition (SVD). The covariance matrix $\mathbf{\Sigma}$ is factorized as $\mathbf{\Sigma} = \mathbf{U}\mathbf{S}\mathbf{U}^{\mathrm T}$, where unitary matrix $\mathbf{U}$ contains singular vectors as columns, and matrix $\mathbf{S}$ has the singular values in the diagonal. Using the $r$ leading singular values, the projection matrix $\mathbf{P}_r$ is defined as
\begin{equation}
\mathbf{P}_r = \mathbf{U}_r \mathbf{S}_r^{1/2} = [\sqrt{s_1}\vec{u}_1, \ldots, \sqrt{s_r}\vec{u}_r] .
\label{Pdef}
\end{equation}
Now it is easy to see that 
\begin{equation}
\Vert \mathbf{P}_r \vec{\alpha}_k  \Vert _{\mathbf{\Sigma}} =  \vec{\alpha}_k^{\mathrm T} \mathbf{P}_r ^{\mathrm T}\mathbf{\Sigma}^{-1} \mathbf{P}_r \vec{\alpha}_k =\vec{\alpha}_k^{\mathrm T}\vec{\alpha}_k. 
\end{equation}
Thus this selection of the projection matrix whitens the prior, and posterior distribution (\ref{P1}) further simplifies to 
\begin{equation}
p( \vec{\alpha}_k\vert \vec{y}_k )  \propto \exp \left( -\frac{1}{2} \left(    \Vert  \vec{y}_k -\mathbf{H}_k (\vec{\mu}_k + \mathbf{P} \vec{\alpha}_k) \Vert^2_{\mathbf{R}_k} + \Vert \vec{\alpha}_k  \Vert   \right)\right) .
\label{P2}
\end{equation}
Now following (\ref{Gauss1}--\ref{Gauss2}) the maximum a posteriori estimate for the parameter vector $\vec{\alpha}_k$ and its error covariance matrix can be obtained from
\begin{eqnarray}
\label{Gauss3}
\vec{\alpha}_k^{\mathrm{est}} & = & \mathbf{\Psi}_k^{\mathrm{est}}(\mathbf{H}_k \mathbf{P}_r)^{\mathrm T}\mathbf{R}_k^{-1}(\vec{y}_k-\mathbf{H}_k \vec{\mu}_k), \\ 
\mathbf{\Psi}_k^{\mathrm{est}} & = & ((\mathbf{H}_k \mathbf{P}_r)^{\mathrm T}\mathbf{R}_k^{-1}(\mathbf{H}_k \mathbf{P}_r) + \mathbf{I})^{-1}.
\label{Gauss4}
\end{eqnarray}

Although we now work within Bayesian framework, an interesting alternative is to use Tikhonov regularization with the same projection matrix $\mathbf{P}_r$ and $\vec{\mu}_k = \vec{0}$. Now the Tikhonov solution (\ref{Tik1}) can be expressed as
\begin{equation}
\vec{\alpha}_k^{\mathrm{tik}} = \left(\left(\mathbf{H}_k \mathbf{P}_r\right)^{\mathrm T}\left(\mathbf{H}_k \mathbf{P}_r\right) + \left(\mathbf{\Gamma}\mathbf{P}_r\right)^{\mathrm T}\left(\mathbf{\Gamma}\mathbf{P}_r \right) \right)^{-1}(\mathbf{H}_k\mathbf{P}_r)^{\mathrm T}\vec{y}_k.
\label{Tik2}
\end{equation}
We note that this Tikhonov solution only uses the prior covariance matrix  to create the projection matrix, and the covariance matrix in itself is not used in the solution.

{\bf Remark.}  The key idea behind prior-based dimension reduction is that the prior covariance matrix $\mathbf{\Sigma}$ is selected so that only small amount of singular vectors are needed to be considered. In context of X-ray images of the size $N\times N$, one now only needs to invert matrix of the size $r\times r$. If now $r$ can be selected much smaller than $N^2$, the direct evaluation of the above formulas is feasible.

{\bf Remark.}   In order to use solutions (\ref{Gauss3}--\ref{Gauss4}) and (\ref{Tik2}) we do not actually need to compute the Radon matrix $\mathbf{H}_k$ explicitly. Matrix representation of the measurement function $\mathbf{H}_k$ is typically created by applying some built-in Radon function to the columns of the identity matrix of the size $N^2\times N^2$. In dimension reduction, we just need to apply this built-in function $r$ times (instead of $N^2$ times) to the columns of $\mathbf{P}_r$.

\subsection{Prior-based dimension reduction in Kalman filtering}
So far at every time step $k$ we have only considered the observations made at the same time. In linear filtering, one considers the following pair
\begin{eqnarray}
\vec{x}_k & = & \mathbf{M}_k \vec{x}_{k-1} + \vec{\xi}_k, \\ 
\vec{y}_k & = & \mathbf{H}_k \vec{x}_k + \vec{\varepsilon}_k.
\end{eqnarray}
Thus in addition to the standard linear inverse problem (\ref{ip}), we have also the dynamic prediction model $\mathbf{M}_k$ that moves the state $\vec{x}_{x-1}$ from the time step $k-1$ to $k$. Errors terms $\vec{\xi}_k$ and $\vec{\varepsilon}_k$ are typically assumed zero-mean and Gaussian: $\vec{\xi}_k \sim N(\vec{0},\mathbf{Q}_k)$ and $\vec{\varepsilon}_k \sim N(\vec{0},\mathbf{R}_k)$ .

In filtering one has two steps: \textit{prediction} and \textit{update}. In the prediction step the estimate of the mean and its covariance matrix of the previous time step is moved to the next time step's prior. In Kalman filtering (e.g., \cite{SarkkaBook}), one calculates
\begin{eqnarray}
\vec{x}_k^{\mathrm p} & = & \mathbf{M}_k \vec{x}_{k-1}^{\mathrm{est}}, \\ 
\mathbf{C}_k^{\mathrm p} & = & \mathbf{M}_k \mathbf{C}_{k-1}^{\mathrm{est}} \mathbf{M}_k^{\mathrm T} +  \mathbf{Q}_k.
\end{eqnarray}
Now using this prior in the update step, the posterior distribution $p(\vec{x}_k \vert \vec{y}_{1:k})$ can be written as
\begin{equation}
p( \vec{x}_k\vert \vec{y}_k ) \propto \exp \left( -\frac{1}{2} \left(    \Vert  \vec{y}_k -\mathbf{H}_k \vec{x}_k \Vert^2_{\mathbf{R}_k} + \Vert \vec{x}_k- \vec{x}_k^{\mathrm p}  \Vert^2_{\mathbf{C}_k^{\mathrm p}}    \right)\right).
\end{equation}
and the posterior mean and covariance matrix can be obtained from the formulas similar to  (\ref{Gauss1}--\ref{Gauss2}), often written in the so called ``m-form,''  as in many Kalman filtering applications the number of observations is typically smaller than the number of states, see e.g. \cite{Clive}.

In dynamic X-ray tomography, regardless of which way we write the formulas (\ref{Gauss1}--\ref{Gauss2}), the direct application of the Kalman filter become unfeasible even for modest size X-ray images. Next we follow \cite{solonen2016} and discuss how the dimension reduction can be used here.

Let us start by parametrizing our state as
\begin{equation}
\vec{x}_k = \vec{x}_k^{\mathrm p}+ \mathbf{P}_r \vec{\alpha}_k.
\end{equation}
Now the prior mean and the covariance matrix in Kalman filter prediction step can be written as
\begin{eqnarray}
\vec{x}_k^{\mathrm p} & = & \mathbf{M}_k (\vec{x}_{k-1}^{\mathrm p}+ \mathbf{P}_r \vec{\alpha}_{k-1}^{\mathrm{est}} ), \\ 
\mathbf{C}_k^{\mathrm p} & = & (\mathbf{M}_k \mathbf{P}_r )(\mathbf{\Psi}_{k-1}^{\mathrm{est}} )(\mathbf{M}_k \mathbf{P}_r )^{\mathrm T} +  \mathbf{Q}_k,
\end{eqnarray}
and the posterior distribution for the parameters $\vec{\alpha}_k$ comes
\begin{equation}
p( \vec{\alpha}_k\vert \vec{y}_k ) \propto \exp \left( -\frac{1}{2} \left(    \Vert  \vec{y}_k -\mathbf{H}_k \vec{x}_k^{\mathrm p} -\mathbf{H}_k \mathbf{P}_r \vec{\alpha}_k \Vert^2_{\mathbf{R}_k} + \Vert  \mathbf{P}_r \vec{\alpha}_k  \Vert^2_{\mathbf{C}_k^{\mathrm p}}    \right)\right).
\end{equation}
The posterior mean and covariance can be obtained from
\begin{eqnarray}
\label{Gauss5}
\vec{\alpha}_k^{\mathrm{est}} & = & \mathbf{\Psi}_k^{\mathrm{est}}(\mathbf{H}_k \mathbf{P}_r)^{\mathrm T}\mathbf{R}_k^{-1}(\vec{y}_k-\mathbf{H}_k \vec{x}_k^{\mathrm p}), \\ 
\mathbf{\Psi}_k^{\mathrm{est}} & = & \left(\left(\mathbf{H}_k \mathbf{P}_r\right)^{\mathrm T}\mathbf{R}_k^{-1}\left(\mathbf{H}_k \mathbf{P}_r\right) +  \mathbf{P}_r^{\mathrm T}(\mathbf{C}_k^{\mathrm p})^{-1}\mathbf{P}_r\right)^{-1}.
\label{Gauss6}
\end{eqnarray}

In principle, the above formulas can be applied as is, however  the direct computation of $(\mathbf{C}_k^{\mathrm p})^{-1}\mathbf{P}_r$ in (\ref{Gauss6}) is not (often) feasible. In order to efficiently compute  $(\mathbf{C}_k^{\mathrm p})^{-1}\mathbf{P}_r$, we write 
\begin{equation}
\mathbf{C}_k^{\mathrm p}  =  (\mathbf{M}_k \mathbf{P}_r )(\mathbf{\Psi}_{k-1}^{\mathrm{est}} )(\mathbf{M}_k \mathbf{P}_r )^{\mathrm T} +  \mathbf{Q}_k = \mathbf{B}_k\mathbf{B}_k^{\mathrm T} +  \mathbf{Q}_k,
\end{equation}
where $\mathbf{B}_k = \mathbf{M}_k \mathbf{P}_r \mathbf{A}_k$, and $\mathbf{A}_k$ is the matrix square root $\mathbf{\Psi}_{k-1}^{\mathrm{est}} = \mathbf{A}_k\mathbf{A}_k^{\mathrm T}$. Now, applying the Sherman-Morrison-Woodbury (SMW) matrix inversion lemma yields
\begin{equation}
(\mathbf{C}_k^{\mathrm p})^{-1}\mathbf{P}_r  = \mathbf{Q}_k^{-1}\mathbf{P}_r-\mathbf{Q}_k^{-1}\mathbf{B}_k(\mathbf{B}_k^{\mathrm T}\mathbf{Q}_k^{-1}\mathbf{B}_k+\mathbf{I})^{-1}\mathbf{B}_k^{\mathrm T}\mathbf{Q}_k^{-1}\mathbf{P}_r,
\label{SMW}
\end{equation}
which can be evaluated as long as $\mathbf{Q}_k^{-1}\mathbf{P}_r$ is easy to compute. In many Kalman filtering application this is the case, and often the model error covariance matrix $\mathbf{Q}_k$ is selected to be diagonal.

{\bf Remark.}   In many applications,  the prediction model $\mathbf{M}_k$ is a natural part of the system. In dynamic X-ray tomography, however, this is not often the case. If no such model is not naturally available, one straightforward option is to model the transport from one image to another simply with identity $\mathbf{M}_k = \mathbf{I}$. This option, of course, assumes that the underlying motion is small. Another option is based on optical flow methods that try to calculate the motion between two consecutive image frames. For example this option was considered in a recent study \cite{burger2017variational} with variational methods. In filtering context, one could for example calculate the displacement field from images $\vec{x}_{k-1}^{\mathrm{est}}$ and $\vec{x}_k^{\mathrm{est}}$, and then use this field to predict $\vec{x}_{k+1}^{\mathrm p} = \mathbf{M}_{\mathrm{flow}}\vec{x}_k^{\mathrm{est}}$.

{\bf Remark.}   As in static case, where we do not need to calculate $\mathbf{H}_k$ explicitly, we also can circumvent the evaluation of $\mathbf{M}_k$. Again we just need to apply the code-level function of $\mathbf{M}_k$   to the columns of $\mathbf{P}_r$ ($r$ times). 

{\bf Remark.}  In this paper, we have presented the dimension reduction Kalman filter only for linear models. The method can also be applied to non-linear filtering, see, e.g., the original paper \cite{solonen2016} for more complex cases.

\subsubsection{Dimension reduction Rauch--Tung--Striebel smoother}
Kalman filter solves dynamically the probability distribution $p(\vec{x}_k,\vec{y}_{1:k})$ given at time step $k$ all the observation $\vec{y}_{1:k}$. The Rauch--Tung--Striebel smoother \cite{RTS} is a fixed interval smoother that solves the probability distribution $p(\vec{x}_k,\vec{y}_{1:K})$ given all the observation $\vec{y}_{1:K}$ for every $k \in [1,\ldots,K]$. The \textit{forward pass} of the smoother is the same as in Kalman filter, and it can be seen as post-processing tool for the Kalman filtering results.

In the \textit{backward pass}, one starts from the final Kalman filtering estimate and operates via recursion. The smoothed state $\vec{x}_{k-1}^{\mathrm{smooth}} $  in RTS smoother can be obtained from
\begin{equation}
\vec{x}_{k-1}^{\mathrm{smooth}} = \vec{x}_{k-1}^{\mathrm{est}} + \mathbf{C}_{k-1}^{\mathrm{est}} \mathbf{M}_k^{\mathrm T} (\mathbf{C}_k^{\mathrm p})^{-1}(\vec{x}_{k}^{\mathrm{smooth}} -\vec{x}_{k}^{\mathrm{p}}),
\label{smorec}
\end{equation}
and its error covariance matrix $\mathbf{C}_{k-1}^{\mathrm{smooth}}$ from
\begin{equation}
\mathbf{C}_{k-1}^{\mathrm{smooth}} = \mathbf{C}_{k-1}^{\mathrm{est}} + \mathbf{C}_{k-1}^{\mathrm{est}} \mathbf{M}_k^{\mathrm T} (\mathbf{C}_k^{\mathrm p})^{-1}(\mathbf{C}_{k}^{\mathrm{smooth}}-\mathbf{C}_{k}^{\mathrm{p}})(\mathbf{C}_k^{\mathrm p})^{-1} \mathbf{M}_k\mathbf{C}_{k-1}^{\mathrm{est}}.
\label{smocovrec}
\end{equation}

Recursive solution~(\ref{smorec}) can be efficiently evaluated by using the SMW formula for $(\mathbf{C}_k^{\mathrm p})^{-1}$ like in Eq.~(\ref{SMW}), and writing $\mathbf{C}_{k-1}^{\mathrm{est}} =\mathbf{P}_r  \mathbf{\Psi}_{k-1}^{\mathrm{est}}\mathbf{P}_r^{\mathrm T}$.

Writing now also $\mathbf{C}_{k-1}^{\mathrm{smooth}}=\mathbf{P}_r \mathbf{\Psi}_{k-1}^{\mathrm{smooth}}\mathbf{P}_r^{\mathrm T} $ in Eq.~(\ref{smocovrec}) yields
\begin{eqnarray}
\mathbf{P}_r \mathbf{\Psi}_{k-1}^{\mathrm{smooth}}\mathbf{P}_r^{\mathrm T} &=& \mathbf{P}_r \mathbf{\Psi}_{k-1}^{\mathrm{est}}\mathbf{P}_r^{\mathrm T}  + \mathbf{P}_r \mathbf{\Psi}_{k-1}^{\mathrm{est}}\mathbf{P}_r^{\mathrm T} \mathbf{M}_k^{\mathrm T} (\mathbf{C}_k^{\mathrm p})^{-1} \times \\
&\times&(\mathbf{P}_r \mathbf{\Psi}_{k}^{\mathrm{smooth}}\mathbf{P}_r^{\mathrm T}-\mathbf{C}_{k}^{\mathrm{p}}) \times \\
&\times&(\mathbf{C}_k^{\mathrm p})^{-1} \mathbf{M}_k\mathbf{P}_r \mathbf{\Psi}_{k-1}^{\mathrm{est}}\mathbf{P}_r^{\mathrm T},
\end{eqnarray}
and further more
\begin{eqnarray}
\mathbf{\Psi}_{k-1}^{\mathrm{smooth}}&=& \mathbf{\Psi}_{k-1}^{\mathrm{est}}  +  \mathbf{\Psi}_{k-1}^{\mathrm{est}}\mathbf{P}_r^{\mathrm T} \mathbf{M}_k^{\mathrm T} (\mathbf{C}_k^{\mathrm p})^{-1} \times \\
&\times&(\mathbf{P}_r \mathbf{\Psi}_{k}^{\mathrm{smooth}}\mathbf{P}_r^{\mathrm T}-\mathbf{C}_{k}^{\mathrm{p}}) \times \\
&\times&(\mathbf{C}_k^{\mathrm p})^{-1} \mathbf{M}_k\mathbf{P}_r \mathbf{\Psi}_{k-1}^{\mathrm{est}}.
\end{eqnarray}
This can be rearragend to
\begin{eqnarray}
\mathbf{\Psi}_{k-1}^{\mathrm{smooth}}&=& \mathbf{\Psi}_{k-1}^{\mathrm{est}} \\
&+&  \mathbf{\Psi}_{k-1}^{\mathrm{est}}\mathbf{D}^{\mathrm T} \mathbf{P}_r \mathbf{\Psi}_{k}^{\mathrm{smooth}}\mathbf{P}_r^{\mathrm T}\mathbf{D} \mathbf{\Psi}_{k-1}^{\mathrm{est}}\\
&-&\mathbf{\Psi}_{k-1}^{\mathrm{est}}\mathbf{D}^{\mathrm T} \mathbf{M}_k\mathbf{P}_r  \mathbf{\Psi}_{k-1}^{\mathrm{est}},
\end{eqnarray}
where $\mathbf{D} = (\mathbf{C}_k^{\mathrm p})^{-1} (\mathbf{M}_k\mathbf{P}_r)$. This can again be evaluated efficiently using the SMW formula.

\section{Numerical experiments\label{numer}}

Here we discuss how the dimension reduction can be applied to 2D static and  3D dynamic X-ray tomography. The purpose is to highlight different features in different methods. We start by applying the ideas to a static problem and then move to dynamic problems with real and simulated data.

\subsection{Tomographic X-ray data of  middle slice of a walnut}
In order to study how the effect of dimension reduction in static X-ray tomography, we use  the FIPS (Finnish Inverse Problems Society) open-access data set of tomographic X-ray data of a walnut \cite{2015arXiv150204064H}. 

We start by creating a $10000\times10000$ prior covariance matrix for $100\times100$ image using standard Gaussian covariance function
\begin{equation}
\mathbf{\Sigma}_{i,j} = \sigma^2\exp \left( -\frac{d(x_i,x_j)^2}{2l^2} \right), 
\label{covfun}
\end{equation}
where $\sigma^2$ is the variance parameter, $l$ is the correlation length and $d(x_i,x_j)$ is the Euclidean distance between pixels $x_i$ and $x_j$. For practical purposes we select $\sigma = 1$ and $l = 3$. We calculate the SVD for $\mathbf{\Sigma}$, and form $\mathbf{P}_r$ using Eq.~(\ref{Pdef}). Finally, we linearly interpolate each column of $\mathbf{P}_r$ to the target size $328 \times 328$.

We start by using dimension reduction Tikhonov regularization. As a Tikhonvov matrix we set $\mathbf{\Gamma} =\gamma\mathbf{I}$ with $\gamma=10$. Now the formula (\ref{Tik2}) can be written as
\begin{equation}
\vec{\alpha}^{\mathrm{tik}} = \left(\left(\mathbf{H} \mathbf{P}_r\right)^{\mathrm T}\left(\mathbf{H} \mathbf{P}_r\right) +\gamma^2\mathbf{S}_r \right)^{-1}(\mathbf{H}\mathbf{P}_r)^{\mathrm T}\vec{y}.
\label{Tik3}
\end{equation}

\begin{figure}[!t]
\centering
\includegraphics[width=4.5in]{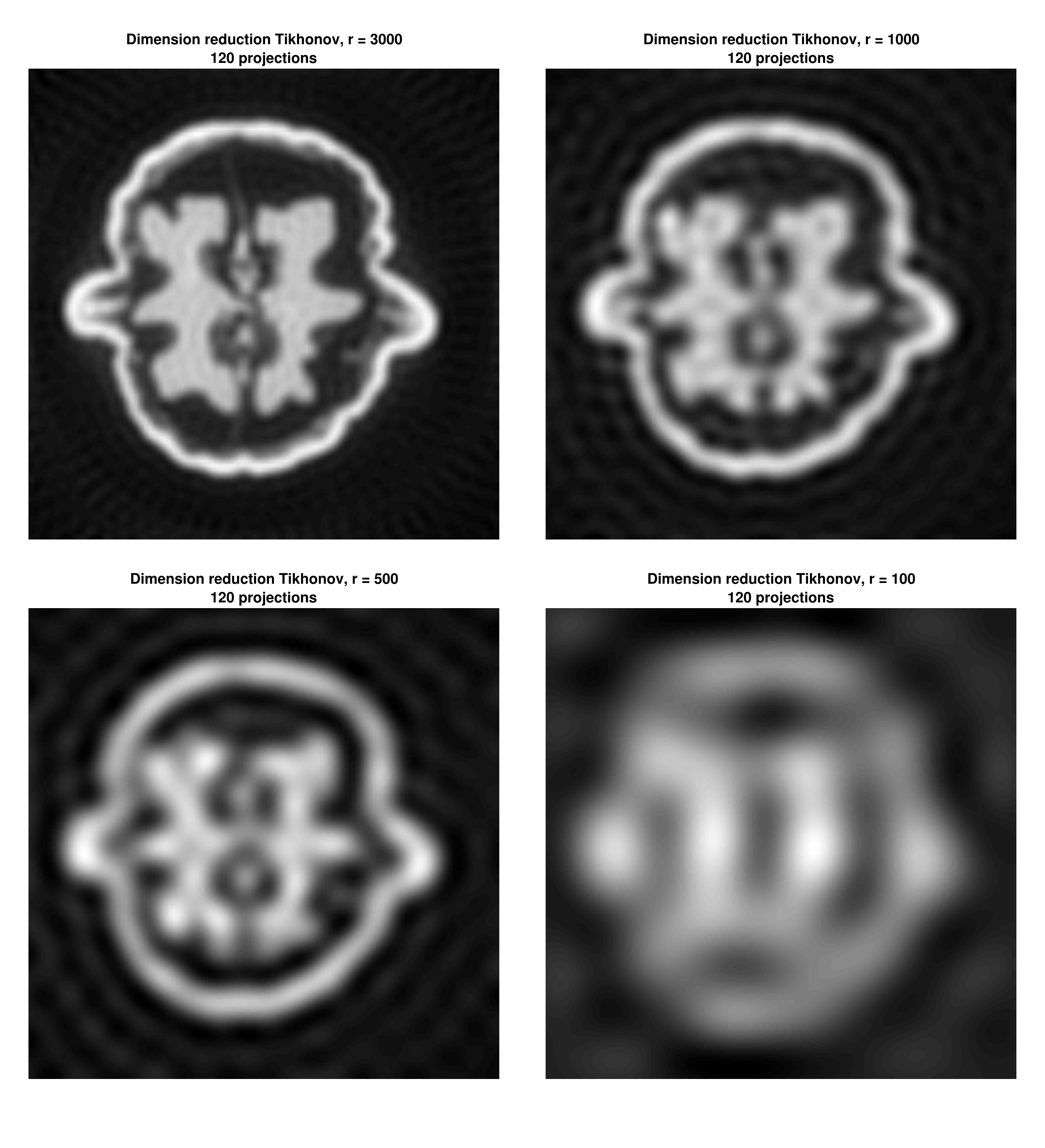}
\caption{Dimension reduction Tikhonov regularization reconstructions of  a walnut using different number of basis vectors ($r = 3000$, 1000, 500 or 100). Open-access FIPS  data set obtained from \protect\url{http://fips.fi/dataset.php}.}
\label{F1}
\end{figure}

Figure~\ref{F1} illustrates results using dimension reduction Tikhonov regularization with different number of basis vectors  ($r = 3000$, 1000, 500 or 100). We can see that lowering the number of basis vectors smoothens the reconstruction. However, the shape of the walnut is clearly recognizable  even with $r = 500$, although the full image size is substantially larger $328^2 = 107584$.

\begin{figure}[!t]
\centering
\includegraphics[width=4.5in]{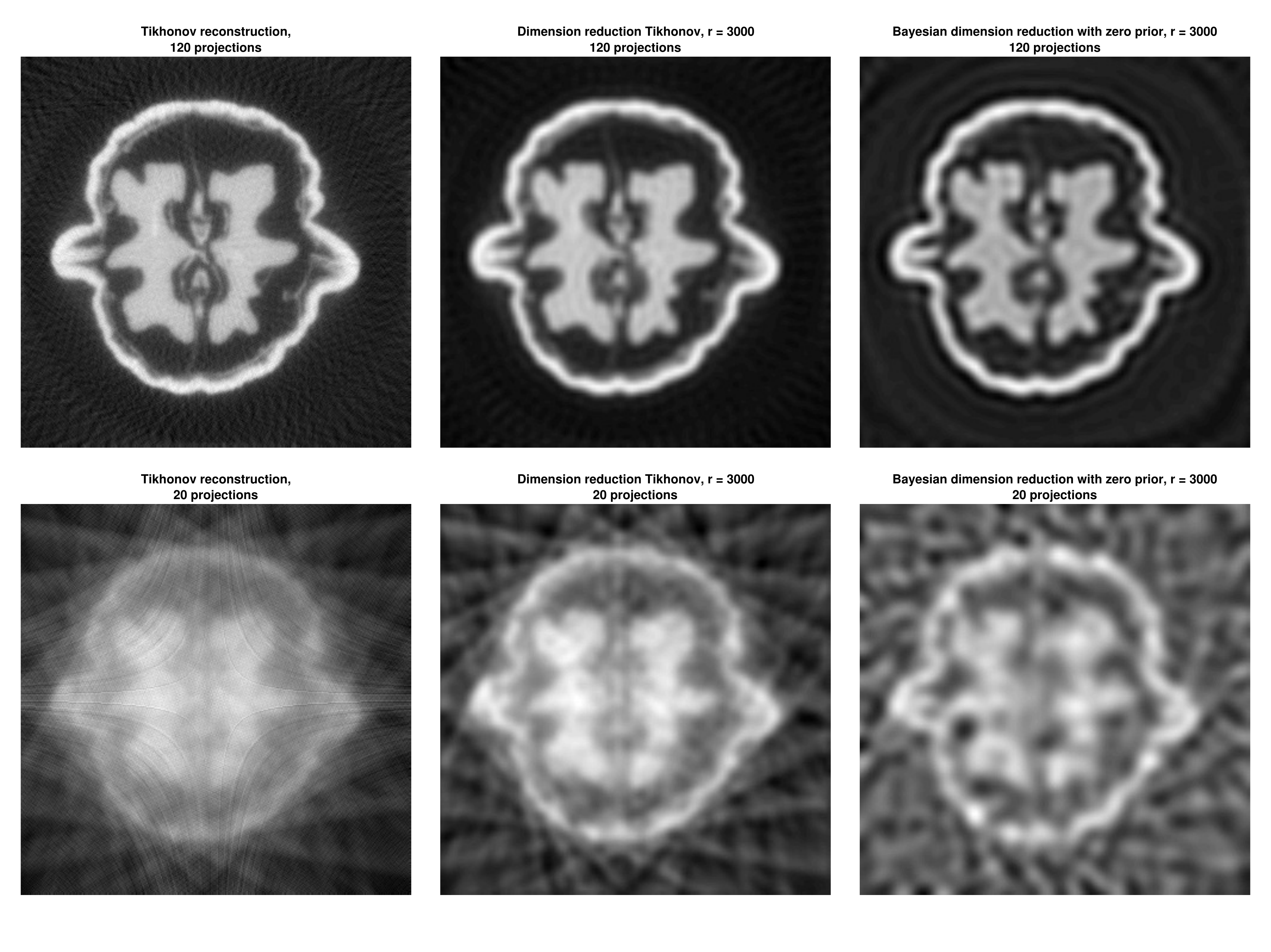}
\caption{Different reconstructions of a walnut with variational Tikhonov regularization (left), dimension reduction ($r = 3000$) Tikhonov regularization (center) and Bayesian dimension reduction (right) with 120 (top) and 20 (bottom) projection angles. Open-access FIPS  data set obtained from \protect\url{http://fips.fi/dataset.php}.}
\label{F2}
\end{figure}

Figure~\ref{F2} illustrates reconstructions of a walnut using variational Tikhonov regularization \cite{2015arXiv150204064H} and dimension reduction Tikhonov regularization ($r = 3000$) with 120 and 20  projection angles. By visually comparing the results, we observe that nearly all the same features are observable, and the results are only slightly smoother with dimension reduction. Moving from 120 to 20 projection angles provides much more drastic change. We also illustrate Bayesian estimate using prior $\vec{\mu} = \vec{0}$. In comparison to Tikhonov solution, we observe some additional artifacts, but the results are still very much acceptable.

%\begin{figure*}[t]
%\vspace*{2mm}
%\begin{center}
%\includegraphics[width=1.0\linewidth] {./F1.eps}
%\end{center}
%\caption{Dimension reduction Tikhonov regularization reconstructions of  a walnut using different number of basis vectors ($r = 3000$, 1000, 500 or 100). Open-access FIPS  data set obtained from \url{http://fips.fi/dataset.php}.}
%\label{F1}
%\end{figure*}

%\begin{figure*}[t]
%\vspace*{2mm}
%\begin{center}
%\includegraphics[width=1.0\linewidth] {./F2.eps}
%\end{center}
%\caption{Different reconstructions of a walnut with variational Tikhonov regularization (left), dimension reduction ($r = 3000$) Tikhonov regularization (center) and Bayesian dimension reduction (right) with 120 (top) and 20 (bottom) projection angles. Open-access FIPS  data set obtained from \url{http://fips.fi/dataset.php}.}
%\label{F2}
%\end{figure*}

\subsection{Dynamic X-ray tomography: stop motion 3D emoji data}
For the real dynamic X-ray dataset we use the 3D emoji dataset measured at the University of Helsinki, see \cite{meaney2018tomographic} for details. The stop motion emoji data is available at 33 time steps with 60 angular directions. The target size is $128 \times 128$.

Again, we use standard Gaussian covariance function (\ref{covfun}) and set $\sigma = 0.1$ and $l = 1.5$. Both observation and model error covariance matrices are set to $\sigma^2\mathbf{I}$. With this rather arbitrary setup, we get visually good results. One option for tuning these parameters, is to use marginal filter likelihood, see, e.g., \cite{npg-21-919-2014} for more discussion. 

As our aim is to minimize the acquisition time, storing big datasets and preferably the radiation dose  (as motivated by sampling {\it in vivo} samples), we perform filtering only with four (rotating) angular directions. This means that in the beginning, the Kalman filter reconstruction is expected to be rather poor, and gradually improve as more data becomes available. Flowing \cite{purisha2017accelerated}, we also enforce non-negativity of the X-ray attenuation and set all negative pixels to zero after each time step. As a ``ground truth reconstruction'' we use static dimension reduction ($r = 1000$) Tikhonov solution from all 60 angular directions. Finally, we post-process the Kalman filtering results using RTS smoother.

\begin{figure}[!t]
\centering
\includegraphics[width=4.5in]{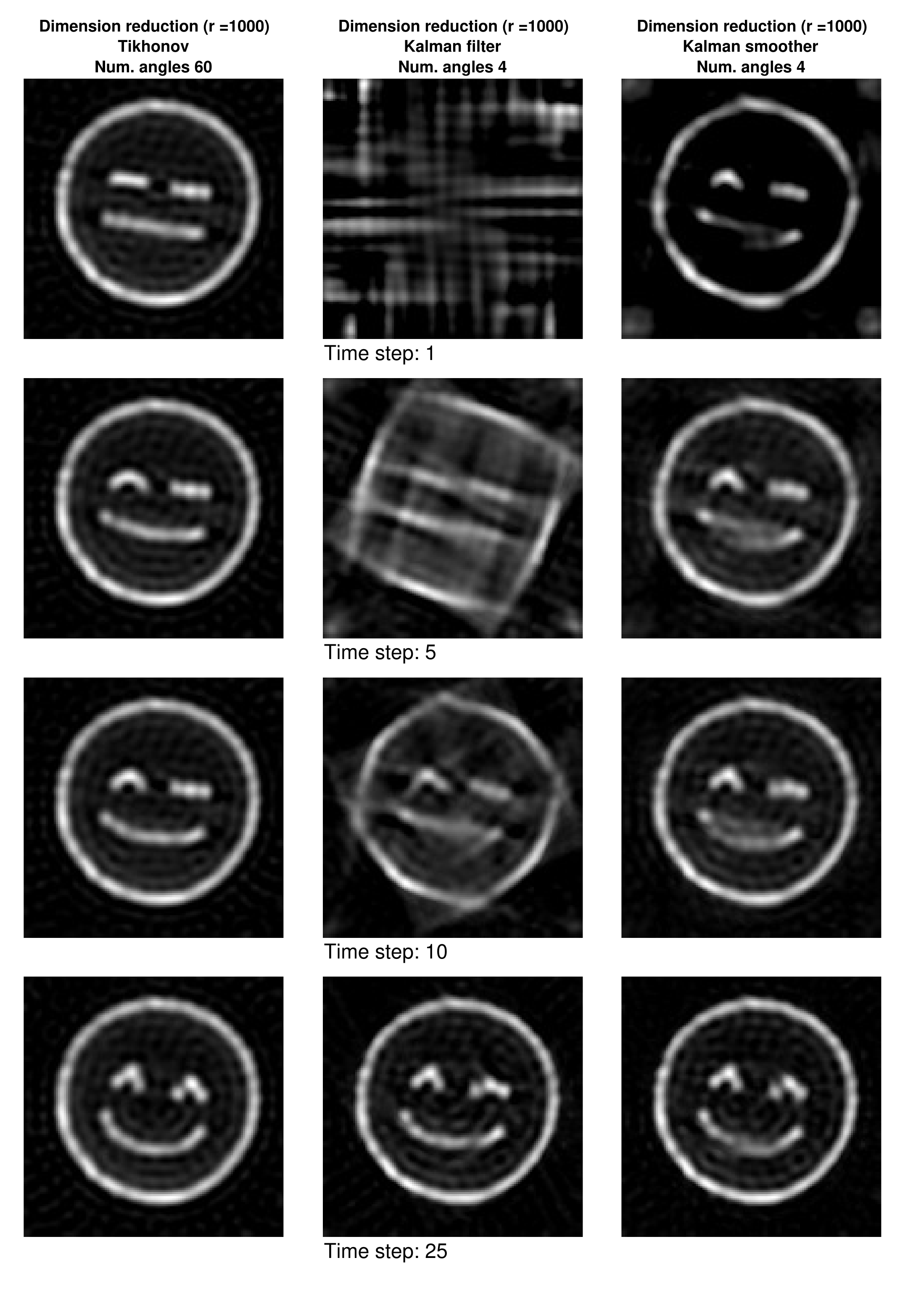}
\caption{Snapshots of stop motion emoji reconstructions at time steps $k=1$, 5, 10 and 25. Dimension reduction ($r = 1000$) methods with Tikhonov regularization (60 angles), Kalman filter (4 angles) and RTS smoother (4 angles). Animation available at \protect\url{https://dl.dropboxusercontent.com/s/159388jcu2mpljm/EmojiAnimation.gif}.}
\label{F3}
\end{figure}

First, we select identity forward model $\mathbf{M}_k = \mathbf{I}$. Figure~\ref{F3} illustrates stop motion emoji results at four time steps $k=1$, 5, 10 and 25. A GIF animation with all 33 time step is available at \url{https://dl.dropboxusercontent.com/s/159388jcu2mpljm/EmojiAnimation.gif}. As expected, in the first steps the Kalman filtering reconstruction is rather poor, but already at step five, the shape of the emoji starts to be visible. The RTS smoother solution is visually closer to the ``ground truth reconstruction,''  although, in the begging the right eye seems to be a little too open. One option could be rerunning the Kalman filter from the RTS smoother initial values. At time step 25, the visual quality of all the reconstructions seem be rather equal, although, Kalman filter seems to be lagging behind (e.g., left eye is little bit too close), because we only use 4 angular directions and identity forward model.

\begin{figure}[!t]
\centering
\includegraphics[width=4.5in]{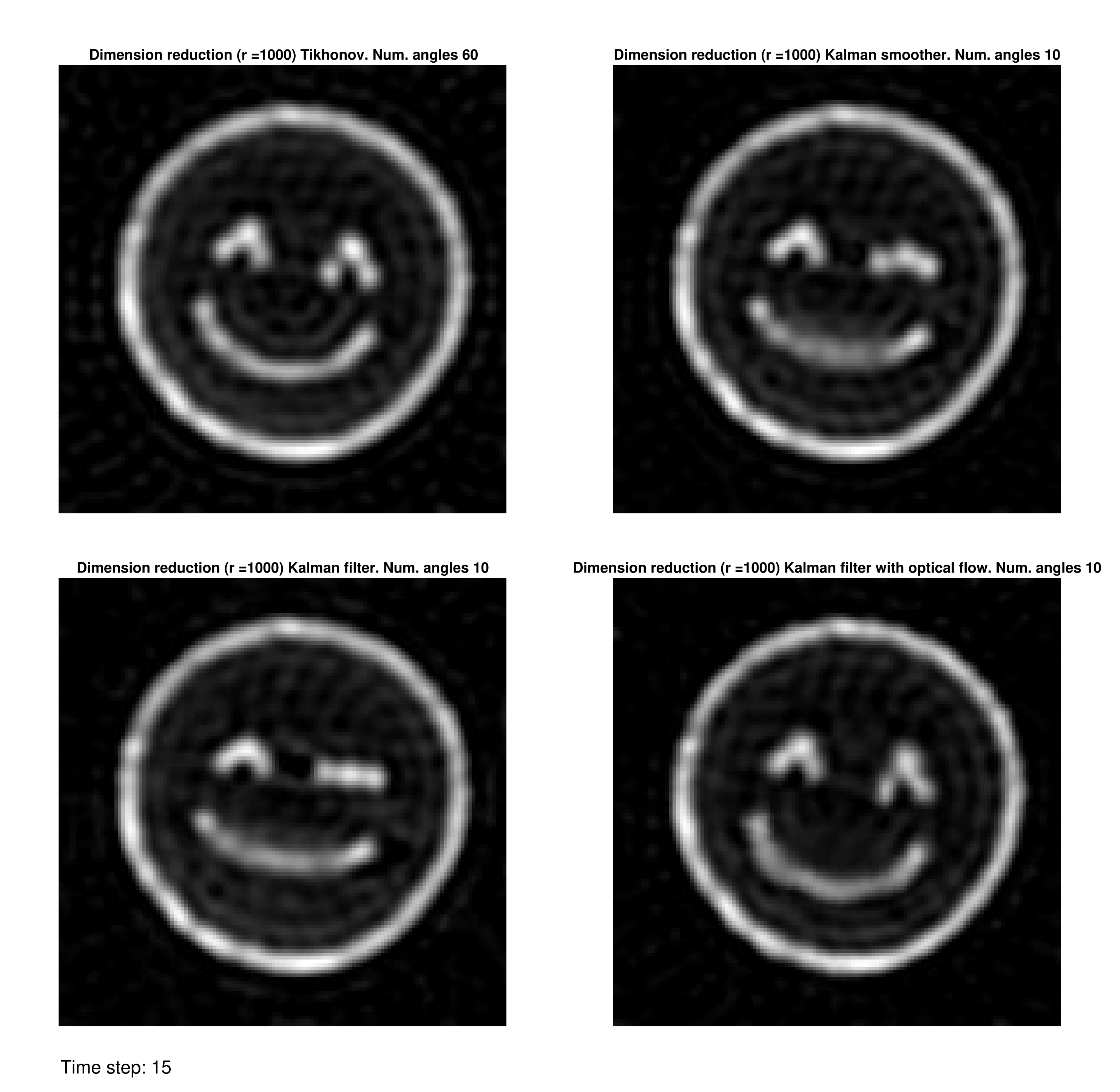}
\caption{Snapshots of stop motion emoji reconstructions at time step $k=15$. Dimension reduction ($r = 1000$) methods with Tikhonov regularization (60 angles), Kalman filter (10 angles) and RTS smoother (10 angles). Now also Kalman filter with optical flow  is illustrated. Note the effect on left eye. Optical flow is calculated using large displacement optical flow Matlab code available at \protect\url{https://www.cs.cmu.edu/~katef/LDOF.html}.}
\label{F4}
\end{figure}

We test this setup also using large displacement optical flow \cite{5551149} forward model, where we compute the displacement field between $\vec{x}_{k-1}^{\mathrm{est}}$ and $\vec{x}_k^{\mathrm{est}}$, and then use this field to predict $\vec{x}_{k+1}^{\mathrm p}$. The idea works technically well, but with just four angular directions  the quality of the images are not good enough for calculating reliable flow, and also rotating angular directions are confused with physical motion. If we increase the number of angular directions to 10, the results with optical flow Kalman filter become more robust. As an example, Figure~\ref{F4} shows a snapshot of this setup at time step 15. Optical flow Kalman filter is started from time step 10. We observe that  the shapes of the eyes and the mouth (left eye, in particular) are now much closer to those of the ground truth reconstruction with optical flow Kalman filter than with the identity version or even with RTS smoother.

%\begin{figure*}[t]
%\vspace*{2mm}
%\begin{center}
%\includegraphics[width=0.8\linewidth] {./F3.eps}
%\end{center}
%\caption{Snapshots of stop motion emoji reconstructions at time steps $k=1$, 5, 10 and 25. Dimension reduction ($r = 1000$) methods with Tikhonov regularization (60 angles), Kalman filter (4 angles) and RTS smoother (4 angles). Animation available at \url{https://dl.dropboxusercontent.com/s/159388jcu2mpljm/EmojiAnimation.gif}.}
%\label{F3}
%\end{figure*}

%\begin{figure*}[t]
%\vspace*{2mm}
%\begin{center}
%\includegraphics[width=1\linewidth] {./F4.eps}
%\end{center}
%\caption{Snapshots of stop motion emoji reconstructions at time step $k=15$. Dimension reduction ($r = 1000$) methods with Tikhonov regularization (60 angles), Kalman filter (10 angles) and RTS smoother (10 angles). Now also Kalman filter with optical flow  is illustrated. Note the effect on left eye. Optical flow is calculated using large displacement optical flow Matlab code available at \url{https://people.eecs.berkeley.edu/~katef/LDOF.html}.}
%\label{F4}
%\end{figure*}

\subsection{Dynamic X-ray tomography: simulated chest phantom}

Finally, in order to test the approach in simulated environment, we use Matlab's 3D~chest~dataset (\url{https://se.mathworks.com/help/images/segment-lungs-from-3-d-chest-mri-data.html}). CT scan has been used to evaluate the blockages, injuries, tumors or other lesions, or other health problems \cite{gohagan2004baseline,puybasset1998computed}. One application is also to determine the pulmonary artery diameter (MPAD) \cite{tan1998utility}, and our proposed method seems to be promising to provide better reconstruction.

In detail, we set our target resolution to $128\times128$, and in order to avoid inverse crime (see, e.g., \cite{SiltanenBook}), we simulate the sinograms with much higher resolution, i.e.,  $512\times512$, and add 1\% noise. Parallel beam geometry is used in the computation. To create data we use 60 equally spaced angles out of 180$^\circ$.

\begin{figure}[!t]
\centering
\includegraphics[width=4.5in]{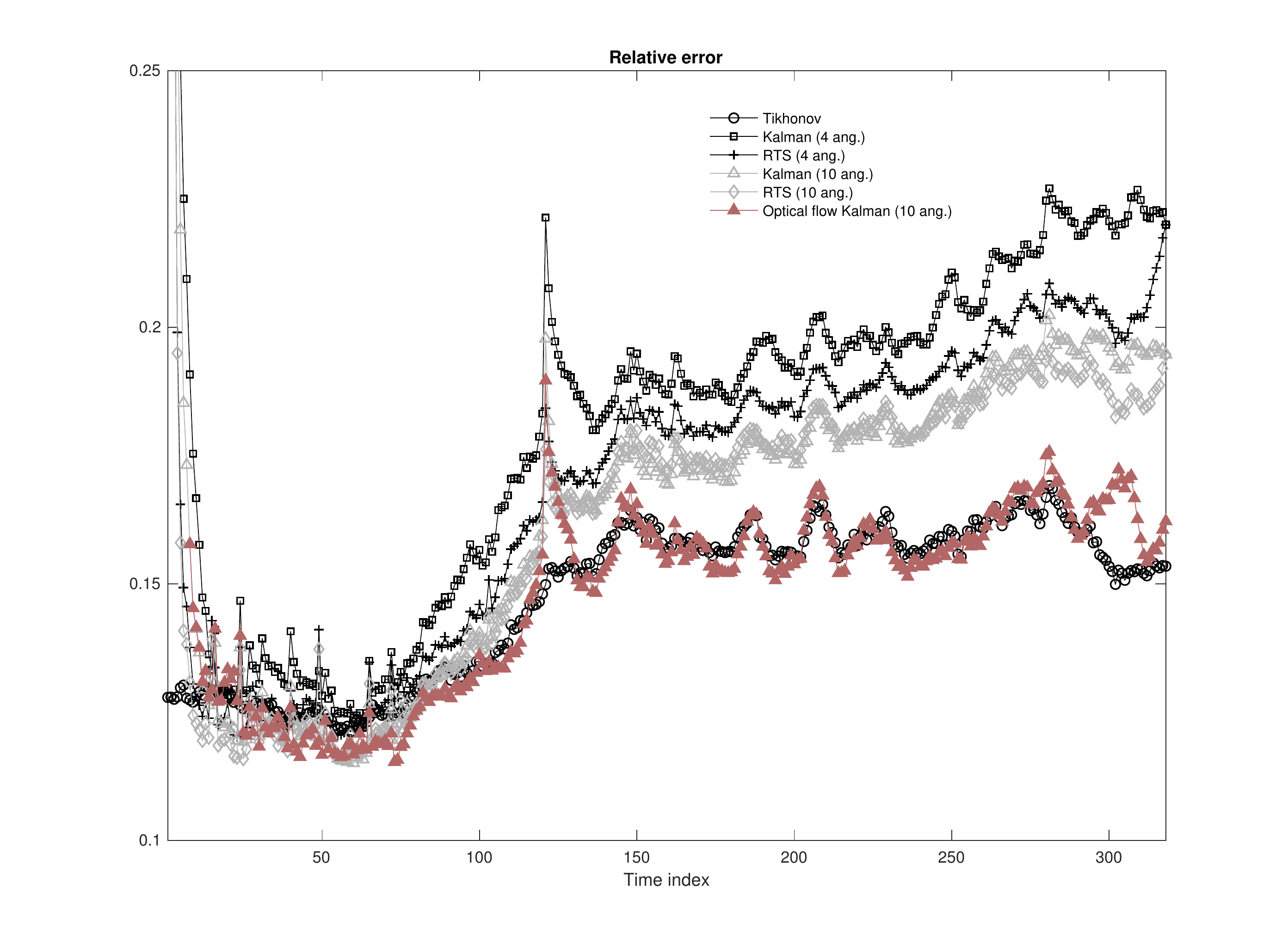}
\caption{Relative errors of dimension reduction ($r = 3000$) methods with Tikhonov regularization (60 angles), Kalman filter (4 and 10 angles), RTS smoother (4 and 10 angles), and optical flow Kalman filter (10 angles).}
\label{F6}
\end{figure}

For Kalman filtering covariance matrices etc., we use the very same setup as with emoji data, and  start the simulations using  identity forward model $\mathbf{M}_k = \mathbf{I}$. An animation of the simulation with four angular directions is available at \url{https://drive.google.com/file/d/1tloIXYErEXeH2WAM9Bywzir0GcB9ILZ6/view?usp=sharing}. Figure~\ref{F6} illustrates relative errors calculated against true phantom.  We can see that after about 15 time steps dimension reduction ($r = 3000$) Kalman filter and RTS smoother yield about the same relative error with four angles as the dimension reduction Tikhonov solution with 60 angles. After time step 75 all relative errors become higher as the shape of the lungs become visible. The relative errors obtained with Kalman filter and RTS smoother with four angular directions are also higher than those with obtained with 60-angle Tikhonov. RTS yields systematically lower relative error than Kalman filter.

\begin{figure}[!t]
\centering
\includegraphics[width=4.5in]{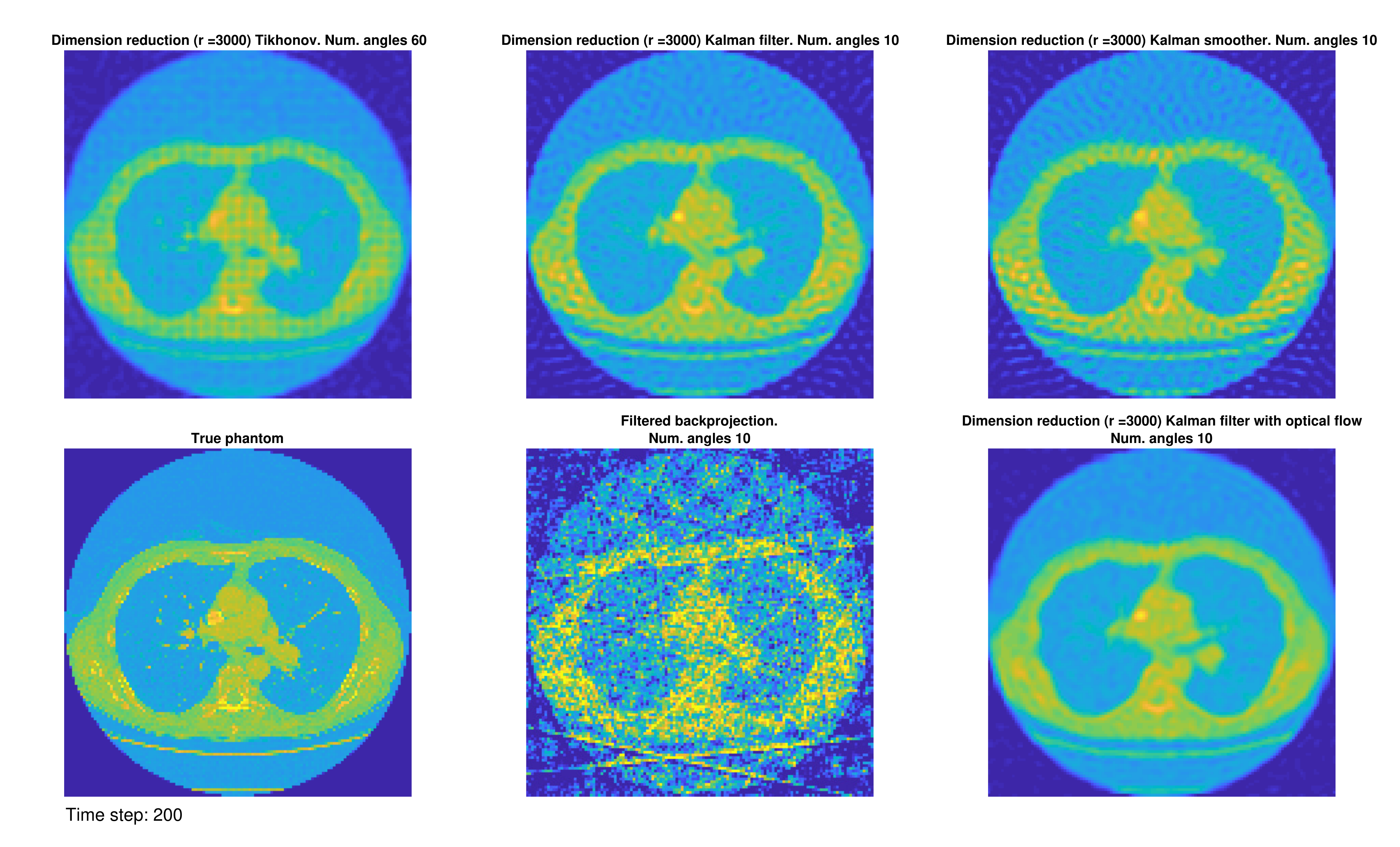}
\caption{Snapshots of 3D~chest~dataset at time step $k=200$. On top row, dimension reduction ($r = 1000$) methods with Tikhonov regularization (60 angles), Kalman filter (10 angles) and RTS smoother (10 angles). On bottom row, the true phantom, FPB reconstruction, and the optical flow Kalman filter reconstruction. Animation available at \protect\url{https://drive.google.com/file/d/1Wt1BfsnmswoKu1jxYkIeAjneIBNKrwO2/view?usp=sharing}} 
\label{F5}
\end{figure}

Like with emoji data, also in this example, just four angular directions are not enough to provide robust results for optical flow Kalman filter. If we increase the number of angular directions to 10, the results with optical flow Kalman filter become robust. An animation with this setup, is available at \url{https://drive.google.com/file/d/1Wt1BfsnmswoKu1jxYkIeAjneIBNKrwO2/view?usp=sharing}, and Figure~\ref{F5} shows a snapshot at time step 200. From the relative errors illustrated in Figure~\ref{F6}, we can note that the relative errors obtained with 10 angular directions are lower than those obtained with four angular directions, as expected. Dimension reduction Kalman filter with optical flow provides the lowest relative errors, and they very close to the the 60-angle Tikhonov solution relative errors.

%\begin{figure*}[t]
%\vspace*{2mm}
%\begin{center}
%\includegraphics[width=1.0\linewidth] {./chest_200_b10_ang.eps}
%\end{center}
%\caption{Snapshots of 3D~chest~dataset at time step $k=200$. On top row, dimension reduction ($r = 1000$) methods with Tikhonov regularization (60 angles), Kalman filter (10 angles) and RTS smoother (10 angles). On bottom row, the true phantom, FPB reconstruction, and the optical flow Kalman filter reconstruction. Animation available at \url{https://drive.google.com/file/d/1Wt1BfsnmswoKu1jxYkIeAjneIBNKrwO2/view?usp=sharing}} 
%\label{F5}
%\end{figure*}

%\begin{figure*}[t]
%\vspace*{2mm}
%\begin{center}
%\includegraphics[width=1.0\linewidth] {./chest_relative_error.eps}
%\end{center}
%\caption{Relative errors of dimension reduction ($r = 3000$) methods with Tikhonov regularization (60 angles), Kalman filter (4 and 10 angles), RTS smoother (4 and 10 angles), and optical flow Kalman filter (10 angles).}
%\label{F6}
%\end{figure*}

\section{Concluding remarks\label{conclusions}}
In recent years, several new methods have been proposed for dynamic X-ray tomography. In this paper, we discussed how the prior-based dimension reduction methods, developed recently for Kalman filtering \cite{solonen2016}, can be applied for the task. The methods discussed here are computationally relatively light, and can be evaluated without variational computations. In Kalman filtering, the reconstructions can be calculated online without solving all the reconstructions at once. The same is true also for the Rauch--Tung--Striebel smoother, as the solution can be calculated recursively after the Kalman filtering has been performed.

In order to give some idea of computation times, we mention that in emoji example, one time step of Matlab implemented  dimension reduction Kalman filter takes about two seconds and one RTS smoother back pass (without calculating the error covariance matrix) about 0.6 seconds (MacBook Pro, 2016, 2.7 GHz Intel Core i7, RAM 16 Gt). The optical flow, calculated using large displacement optical flow Matlab code (\url{https://www.cs.cmu.edu/~katef/LDOF.html}) adds about five seconds per time step.

The numerical results, presented in this paper are promising for future applications.  The dimension reduction Kalman filter with identity forward model, seem to be extremely robust, and provide acceptable results even with just four angular directions. The setup for the filter was also rather arbitrary, and no careful tuning for the filter parameters was needed. The filtering results can be further smoothed with dimension reduction version of the RTS smoother and it improved the quality of the reconstructions especially in the beginning of the simulation. Slightly more complicated version of the dimension reduction Kalman filter can be obtained using optical flow methods. However, the quality of the reconstructions has to be \textit{good enough} in order for it to work properly. For example, we could not get this version of the dimension reduction Kalman filter to work properly with just four angular directions. When the number of angular directions was increased the method became robust, and performed better (no lagging, lower relative error) than the Kalman filter and the RTS smoother with identity forward model.

Future work is to implement the strategy in {\it in vivo} samples or cardiac imaging and to implement in limited angles or random angles tomography problems. Reconstruct a human chest or lung from real measured data using the proposed method would be worth to try as well.
Finally we note that the  dimension reduction filter and smoother performed just as expected, and the main limitation factor was the number of data points at each time step. The dimension reduction in itself did not seem to harm the solution too much. This was even further tested with the static walnut example where dimension reduction method with just 3000 basis vectors provided visually nearly the same quality as the full inversion with the number of unknowns of $328^2 = 107584$. This feature is of course very much case dependent.

\section*{Acknowledgements}
Authors acknowledge support from TEKES/Business Finland ``Heartbeat'' project 6614/31/2016, the Academy of Finland through the Finnish Centre of Excellence in Inverse Problems Research 2012-2017 (decision number 284715), the Finnish Centre of Excellence in Inverse Modelling and Imaging 2018--2025, decision numbers 312119, 312125 and 312122. Janne Hakkarainen thanks Marko Laine for fruitful discussions on practical implementation of dimension reduction RTS smoother.


\begin{thebibliography}{10}

\bibitem{berninger1980multiple}
Walter~H Berninger and Rowland~W Redington.
\newblock Multiple purpose high speed tomographic x-ray scanner, April~1 1980.
\newblock US Patent 4,196,352.

\bibitem{bonnet2003dynamic}
Stephane Bonnet, Anne Koenig, S{\'e}bastien Roux, Patrick Hugonnard, R{\'e}gis
  Guillemaud, and Pierre Grangeat.
\newblock Dynamic {X}-ray computed tomography.
\newblock {\em Proceedings of the IEEE}, 91(10):1574--1587, 2003.

\bibitem{5551149}
T.~Brox and J.~Malik.
\newblock Large displacement optical flow: Descriptor matching in variational
  motion estimation.
\newblock {\em IEEE Transactions on Pattern Analysis and Machine Intelligence},
  33(3):500--513, March 2011.

\bibitem{bubba2017shearlet}
T.~A. Bubba, M.~M{\"a}rz, Z.~Purisha, M.~Lassas, and S.~Siltanen.
\newblock Shearlet-based regularization in sparse dynamic tomography.
\newblock In {\em Wavelets and Sparsity XVII}, volume 10394, page 103940Y.
  International Society for Optics and Photonics, 2017.

\bibitem{burger2017variational}
Martin Burger, Hendrik Dirks, Lena Frerking, Andreas Hauptmann, Tapio Helin,
  and Samuli Siltanen.
\newblock A variational reconstruction method for undersampled dynamic x-ray
  tomography based on physical motion models.
\newblock {\em Inverse Problems}, 33(12):124008, 2017.

\bibitem{buzug2008computed}
Thorsten~M. Buzug.
\newblock {\em Computed tomography: from photon statistics to modern cone-beam
  {CT}}.
\newblock Springer Science \& Business Media, 2008.

\bibitem{gohagan2004baseline}
John Gohagan, Pamela Marcus, Richard Fagerstrom, Paul Pinsky, Barnett Kramer,
  and Philip Prorok.
\newblock Baseline findings of a randomized feasibility trial of lung cancer
  screening with spiral {CT} scan vs chest radiograph: the lung screening study
  of the {National Cancer Institute}.
\newblock {\em Chest}, 126(1):114--121, 2004.

\bibitem{hahn2014reconstruction}
Bernadette Hahn.
\newblock Reconstruction of dynamic objects with affine deformations in
  computerized tomography.
\newblock {\em Journal of Inverse and Ill-posed Problems}, 22(3):323--339,
  2014.

\bibitem{hahn2014efficient}
Bernadette~N Hahn.
\newblock Efficient algorithms for linear dynamic inverse problems with known
  motion.
\newblock {\em Inverse Problems}, 30(3):035008, 2014.

\bibitem{hahn2015dynamic}
Bernadette~N. Hahn.
\newblock Dynamic linear inverse problems with moderate movements of the
  object: Ill-posedness and regularization.
\newblock {\em Inverse Problems \& Imaging}, 9(2), 2015.

\bibitem{2015arXiv150204064H}
K.~{H{\"a}m{\"a}l{\"a}inen}, L.~{Harhanen}, A.~{Kallonen},
  A.~{Kujanp{\"a}{\"a}}, E.~{Niemi}, and S.~{Siltanen}.
\newblock {Tomographic X-ray data of a walnut}.
\newblock {\em ArXiv e-prints}, February 2015.

\bibitem{avinash1988principles}
Avinash~C. Kak and Malcolm Slaney.
\newblock {\em Principles of computerized tomographic imaging}.
\newblock IEEE press, 1988.

\bibitem{Katsevich2010}
A.~Katsevich.
\newblock An accurate approximate algorithm for motion compensation in
  two-dimensional tomography.
\newblock {\em Inverse Problems}, 26(6):065007, 16, 2010.

\bibitem{liu2001half}
Ying Liu, Hong Liu, Ying Wang, and Ge~Wang.
\newblock {Half-scan cone-beam CT fluoroscopy with multiple x-ray sources}.
\newblock {\em Medical physics}, 28(7):1466--1471, 2001.

\bibitem{Marzouk20091862}
Youssef~M. Marzouk and Habib~N. Najm.
\newblock Dimensionality reduction and polynomial chaos acceleration of
  {B}ayesian inference in inverse problems.
\newblock {\em J. Comput. Phys.}, 228(6):1862--1902, 2009.

\bibitem{meaney2018tomographic}
Alexander Meaney, Zenith Purisha, and Samuli Siltanen.
\newblock Tomographic {X}-ray data of {3D} emoji.
\newblock {\em arXiv preprint arXiv:1802.09397}, 2018.

\bibitem{SiltanenBook}
Jennifer~L. Mueller and Samuli Siltanen.
\newblock {\em Linear and Nonlinear Inverse Problems with Practical
  Applications}.
\newblock SIAM, Philadelphia, 2012.

\bibitem{natterer1986mathematics}
Frank Natterer.
\newblock {\em The mathematics of computerized tomography}, volume~32.
\newblock Siam, 1986.

\bibitem{natterer2001mathematical}
Frank Natterer and Frank W{\"u}bbeling.
\newblock {\em Mathematical methods in image reconstruction}, volume~5.
\newblock Siam, 2001.

\bibitem{niemi2015dynamic}
Esa Niemi, Matti Lassas, Aki Kallonen, Lauri Harhanen, Keijo
  H{\"a}m{\"a}l{\"a}inen, and Samuli Siltanen.
\newblock Dynamic multi-source {X}-ray tomography using a spacetime level set
  method.
\newblock {\em Journal of Computational Physics}, 291:218--237, 2015.

\bibitem{purisha2017accelerated}
Zenith Purisha, Sakari~S. Karhula, Juuso Ketola, Juho Rimpel{\"a}inen, Miika~T.
  Nieminen, Simo Saarakkala, Heikki Kr{\"o}ger, and Samuli Siltanen.
\newblock An automatic regularization method: {A}n application for {3D} {X}-ray
  micro-{CT} reconstruction using sparse data.
\newblock {\em arXiv preprint arXiv:1708.02067v2}, 2017.

\bibitem{puybasset1998computed}
Louis Puybasset, Philippe Cluzel, Nan Chao, Arthur~S. Slutsky, Pierre Coriat,
  Jean-Jacques Rouby, and CT~Scan ARDS~Study Group.
\newblock A computed tomography scan assessment of regional lung volume in
  acute lung injury.
\newblock {\em American journal of respiratory and critical care medicine},
  158(5):1644--1655, 1998.

\bibitem{RTS}
H.~E. Rauch, C.~T. Striebel, and F.~Tung.
\newblock Maximum likelihood estimates of linear dynamic systems.
\newblock {\em AIAA Journal}, 3(8):1445--1450, 2017/08/18 1965.

\bibitem{robb1983high}
Richard~A. Robb, Eric~A. Hoffman, Lawrence~J. Sinak, Lowell~D. Harris, and Erik~L.
  Ritman.
\newblock High-speed three-dimensional x-ray computed tomography: The dynamic
  spatial reconstructor.
\newblock {\em Proceedings of the IEEE}, 71(3):308--319, 1983.

\bibitem{Clive}
Clive~D. Rodgers.
\newblock {\em Inverse methods for atmospheric sounding: theory and practice}.
\newblock World Scientific, 2000.

\bibitem{Roux2004}
S.~Roux, L.~Desbat, A.~Koenig, and P.~Grangea.
\newblock {Exact reconstruction in 2D dynamic CT: compensation of
  time-dependent affine deformations}.
\newblock {\em Physics in Medicine and Biology}, 2004.

\bibitem{SarkkaBook}
Simo S{\"a}rkk{\"a}.
\newblock {\em Bayesian Filtering and Smoothing}.
\newblock Cambridge University Press, 2013.

\bibitem{npg-21-919-2014}
A.~Solonen, J.~Hakkarainen, A.~Ilin, M.~Abbas, and A.~Bibov.
\newblock Estimating model error covariance matrix parameters in extended
  {K}alman filtering.
\newblock {\em Nonlinear Processes in Geophysics}, 21(5):919--927, 2014.

\bibitem{solonen2016}
Antti Solonen, Tiangang Cui, Janne Hakkarainen, and Youssef Marzouk.
\newblock On dimension reduction in {G}aussian filters.
\newblock {\em Inverse Problems}, 32(4):045003, 2016.

\bibitem{stiel1993digital}
Georg~M. Stiel, Ludmilla~S.~G. Stiel, Erhard Klotz, and Christoph~A. Nienaber.
\newblock Digital flashing tomosynthesis: a promising technique for
  angiocardiographic screening.
\newblock {\em IEEE transactions on medical imaging}, 12(2):314--321, 1993.

\bibitem{tan1998utility}
Rana~Teresa Tan, Ronald Kuzo, Lawrence~R. Goodman, Ronald Siegel, George~R.
  Haasler, and Kenneth~W. Presberg.
\newblock {Utility of CT scan evaluation for predicting pulmonary hypertension
  in patients with parenchymal lung disease}.
\newblock {\em Chest}, 113(5):1250--1256, 1998.

\end{thebibliography}
\end{document}